\documentclass[aps,prl,twocolumn,showpacs,groupedaddress,floatfix]{revtex4}  
\usepackage{graphicx}  
\usepackage{dcolumn}   
\usepackage{bm}        
\usepackage{amssymb}   
\usepackage{multirow}
\begin{document}
\leftline{}
\rightline{\mbox{FERMILAB-PUB-10-008-E}}
\title{Dependence of the $\bf t\bar{t}$ production cross section on the \\ transverse momentum of the top quark }
%
\author{V.M.~Abazov$^{37}$}
\author{B.~Abbott$^{75}$}
\author{M.~Abolins$^{64}$}
\author{B.S.~Acharya$^{30}$}
\author{M.~Adams$^{50}$}
\author{T.~Adams$^{48}$}
\author{E.~Aguilo$^{6}$}
\author{G.D.~Alexeev$^{37}$}
\author{G.~Alkhazov$^{41}$}
\author{A.~Alton$^{64,a}$}
% ...

\affiliation{\vspace{0.1 in}(The D\O\ Collaboration)\vspace{0.1 in}}
% ...
\affiliation{$^{6}$Simon Fraser University, Burnaby, British Columbia, Canada;
                and York University, Toronto, Ontario, Canada}
% ...
\affiliation{$^{30}$Tata Institute of Fundamental Research, Mumbai, India}
% ...
\affiliation{$^{37}$Joint Institute for Nuclear Research, Dubna, Russia}
% ...
\affiliation{$^{41}$Petersburg Nuclear Physics Institute,
                St. Petersburg, Russia}
% ...
\affiliation{$^{48}$Florida State University, Tallahassee, Florida 32306, USA}
% ...
\affiliation{$^{50}$University of Illinois at Chicago,
                Chicago, Illinois 60607, USA}
% ...
\affiliation{$^{64}$Michigan State University,
                East Lansing, Michigan 48824, USA}
% ...
\affiliation{$^{75}$University of Oklahoma, Norman, Oklahoma 73019, USA}
% ...
\date{January 12th 2010}

\begin{abstract}
  We present a measurement of the differential cross section for
  $t\bar{t}$ events produced in $p\bar{p}$ collisions at
  $\sqrt{s}=1.96$~TeV as a function of the transverse momentum ($p_T$)
  of the top quark.
  The selected events contain a high-$p_T$ lepton ($\ell$), a large 
  imbalance in $p_T$, four or more jets with at least one candidate
  for a $b$ jet, and correspond to
  1~fb${}^{-1}$ of integrated luminosity recorded with the D0
  detector. 
  Objects in the event are associated through a constrained kinematic
  fit to the $t\bar{t}\rightarrow WbW\bar{b} \rightarrow \ell\nu b \,
  q\bar{q}'\bar{b}$ process.
  Results from next and next-to-next-to-leading-order perturbative QCD calculations
  agree with the measured differential cross section.
  Comparisons are also provided to predictions from Monte Carlo event
  generators using QCD calculations at different levels of precision.
\end{abstract}

\maketitle 

The transverse momentum ($p_T$) of top quarks in $t\bar{t}$ events
provides a unique window on heavy-quark production at large momentum
scales. In the standard model (SM), the lifetime of the top quark is
far shorter than the characteristic hadron-formation time of quantum
chromodynamics (QCD), which provides access to the properties and
kinematics of a ``bare'' quark, such as mass, charge, spin, and $p_T$,
that are almost unaffected by bound-state formation or final-state
interactions~\cite{theory}. 
The top quark is unique in that it has a mass close to the scale of
electroweak symmetry breaking. Detailed studies of the properties
of this bare quark beyond the measurement of its total production rate,
such as the measurement of its quantum numbers and of its couplings
to other SM particles, may indicate whether the top quark plays a
privileged role in the symmetry breaking. 
Focusing on details of the $t\bar{t}$ production, measurements of differential cross
sections in the $t\bar{t}$ system test perturbative QCD (pQCD) for
heavy-quark production and can constrain potential new physics beyond the
SM~\cite{BSM_physics}, e.g., by measuring the transverse momentum 
of the top quark~\cite{top_anomalous}. 

In this Letter, we present a new measurement of the
inclusive differential cross section for $p\bar{p}\rightarrow
t\bar{t}+X$ production at $\sqrt{s}=1.96$~TeV as a function of the
$p_T$ of the top quark. The measurement is corrected for 
detector efficiency, acceptance and resolution effects, making it
possible to perform direct comparisons with different theoretical
predictions. The data were acquired with the D0 detector
at the Fermilab Tevatron Collider and correspond to an integrated
luminosity of $\approx 1$~fb${}^{-1}$. 
This measurement was performed
in the $\ell+$jets decay channel of $t\bar{t}\rightarrow
WbW\bar{b}\rightarrow \ell\nu+b\bar{b}\,+\geq2$~jets, where $\ell$
represents an $e$ or $\mu$ from the decay of the $W$ boson or from
$W\rightarrow\tau\rightarrow\ell$. The dependence of the cross section on the
$p_T$ of the top quark was examined previously using $\approx$
$100\,{\rm pb}^{-1}$ of Tevatron Run~I data at
$\sqrt{s}=1.8$~TeV~\cite{cdf_run1_toppt_d0_run1_topmass_pt}, where no
deviations from the SM were reported.

The D0 detector~\cite{d0det} is equipped with a $2\,$T solenoidal
magnet surrounding silicon-microstrip and scintillating-fiber
trackers.  These are followed by electromagnetic (EM) and hadronic
uranium/liquid argon calorimeters, and a muon spectrometer consisting
of $1.8$~T iron toroidal magnets and wire chambers and scintillation
counters.  Electrons are identified as track-matched energy clusters
in the EM calorimeter.  Muons are identified by matching tracks in the
inner tracking detector with those in the muon spectrometer.  Jets are
reconstructed from calorimeter energies using the Run~II iterative
seed-based midpoint cone algorithm with a radius of 0.5~\cite{d0jets}.
Jets are identified as originating from a $b$ quark using an
artificial neural network ($b$~NN) which combines several tracking
variables~\cite{scanlon}.  Large missing transverse energy,
${\not\!\!E_T}$ (the negative of the vector sum of transverse energies
of calorimeter cells, corrected for reconstructed muons) signifies the
presence of an energetic neutrino.  Events are selected using a
three-level trigger system, which has access to tracking, calorimeter,
and muon information, and assures that only events with the desired
topology or with objects above certain energy thresholds are kept for
further analysis.

The analysis uses similar data
samples, event selection, and corrections as used in the inclusive
$t\bar{t}\rightarrow\ell+$jets cross-section measurements detailed in
Ref.~\cite{xsect_ref}. Events accepted by lepton+jets triggers are subject to additional
selection criteria including exactly one isolated lepton with
$p_T>20$~GeV$/c$ and $\geq4$ jets with $p_T > 20$~GeV$/c$ and
$|\eta|<2.5$~\cite{eta}; at least one jet must have $p_T > 40$
GeV$/c$.  At least one jet is also required to be tagged by the $b$~NN
algorithm.  Additionally, we require ${\not\!\!E_T}>20$~GeV ($25$~GeV)
for the $e+$jets ($\mu+$jets) channel and electrons (muons) with
$|\eta|<1.1$ ($2.0$).

Our measurement uses the {\sc alpgen}~\cite{alpgen} event generator,
with {\sc pythia}~\cite{pythia} for parton showering, hadronization,
and modeling of the underlying event, to simulate the inclusive
$t\bar{t}$ signal.  A {\sc pythia} sample serves as a cross check.
The CTEQ6L1 set of parton distribution functions (PDFs)~\cite{cteq} was used 
with a common factorization and renormalization scale set to 
$\mu = m_{\rm t} + \sum {p_T^{\rm jets}}$ for
$m_{\rm t} = 170\,{\rm GeV}/c^2$.
Backgrounds are modeled with {\sc alpgen+pythia} for $W+$jets and
$Z+$jets production, {\sc pythia} for diboson ($WW$, $WZ$, and $ZZ$)
production, and {\sc comphep}~\cite{Boos:2006af} for single top-quark
production. The detector response is simulated using {\sc
  geant}~\cite{geant}.  The simulated $t\bar{t}$ signal is normalized
to the cross section measured by a dedicated likelihood fit in the 
same final state using the same event selections (including the $b$-tagging 
requirement) and data as Ref.~\cite{xsect_ref}, namely to $8.46^{+1.09}_{-0.97}$~pb 
at a top-quark mass $m_t=170\,{\rm GeV}/c^2$ (in good agreement with the value
extracted in this study by integrating the differential cross section). 
The diboson and single top-quark
backgrounds are normalized to their SM predictions, $Z+$jets to the
prediction from next-to-leading-order (NLO) pQCD, and $W+$jets such
that the predicted number of events matches the data before applying
$b$ tagging.  

\begin{table}[b]
\caption{Expected yields for signal and backgrounds samples and observed event counts in $e$+jets and $\mu+$jets channels.}
\label{table_yields}
\begin{tabular}{l|rr}
\hline \hline
sample &  {$e$+jets} &  {$\mu$+jets} \\ \hline
$t\bar{t}$  & 131& 108 \\
$W+$jets    & 10 & 15 \\
$Z+$jets    & 3.0 & 3.1 \\
single top          & 2.7 & 2.0 \\
${\rm diboson  } $  & 1.3 & 1.3 \\
${\rm multijet } $  & 9.0 & 0.0 \\ \hline
summed prediction   & 156 & 130 \\
total background uncertainty & 3.0 & 2.8 \\
predicted signal uncertainty & 11 & 9.0 \\ \hline
${\rm data     } $  & 145 & 141
\\ \hline \hline
\end{tabular}
\end{table}

The small multijet background, in which a jet is misidentified as an isolated lepton, is 
non-negligible only in the $e$+jets channel.  Its rate is estimated from data using 
the large difference in the probability of electromagnetic showers of real electrons 
or misidentified jets to satisfy the electron selection criteria.
The details of the sample composition and the observed yields before and after 
requiring the jets to be tagged as $b$-jet are presented in Table~\ref{table_yields}.

%
\begin{figure}[t]
\includegraphics[width=0.5\textwidth,trim=10 10 10 20,clip=true]{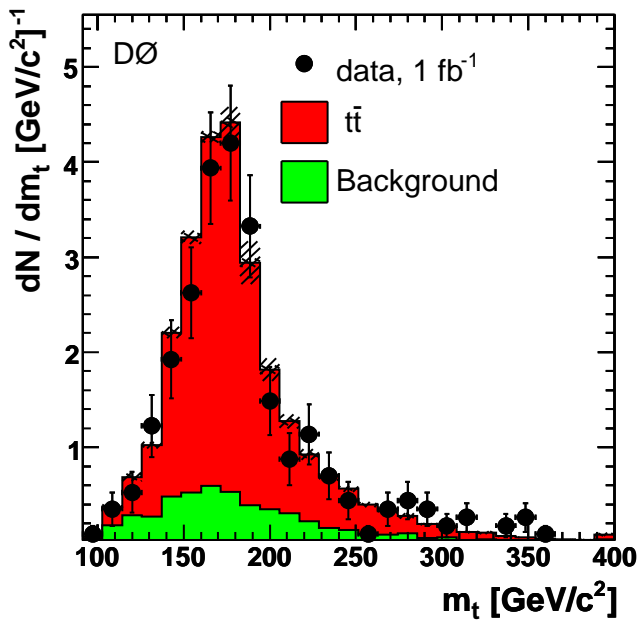}
\caption{The reconstructed top-quark mass compared with
  expectation. Hashed areas represent statistical and jet energy
  calibration uncertainties on the prediction.}
\label{fig:figure1}
\end{figure}
The selection yields 145 and 141 events in the $e+$jets and $\mu+$jets
decay channels, respectively.  The measured $t\bar{t}$ signal fraction
is $0.79$, indicating that this sample is suitable for detailed
studies of $t\bar{t}$ production. A constrained kinematic fit to the
$t\bar{t}$ final state, which takes into account the unreconstructed
neutrino and finite experimental resolution, is used to associate
leptons and jets with individual top
quarks~\cite{hitfit,hitfit_scott}.  The fit assumes equal masses for
the two reconstructed top quarks and the two reconstructed $W$ boson
masses are constrained to~$80.4$~GeV$/c^2$.  All possible permutations
of objects needed to produce the $t\bar{t}$ system are considered, and
the solution of fitted leptonic and hadronic top-quark four-momenta
with the smallest $\chi^2$ (the goodness of the fit) is selected for
further analysis.  
The $b$-jet assignment information is not used in the selection of the
best permutation to avoid the associated efficiency loss. The effects
of possibly selecting a wrong permutation when choosing the one with
the best $\chi^2$ are taken into account in the corrections of the
measurement to the parton level.
The solution with the best (second best) $\chi^2$ corresponds to
the correct assignment of the quarks from the decay of the $t\bar{t}$ pair
in 48\% (17\%) of events.

\begin{figure}[t]
\includegraphics[width=0.50\textwidth,height=0.5\textwidth,trim=10 10 10 20,clip=true]{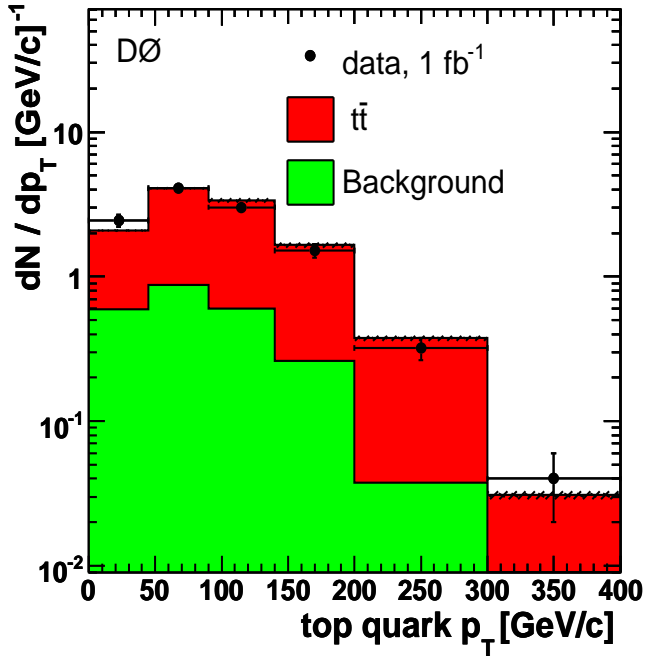}
\caption{The $p_T$ spectrum of top quarks (two entries per event)
  compared with expectation. Hashed areas represent statistical and
  jet energy calibration uncertainties on the prediction.}
\label{fig:figure2}
\end{figure}

\begin{figure}[t]
\includegraphics[width=0.50\textwidth,height=0.5\textwidth,trim=10 10 10 20,clip=true]{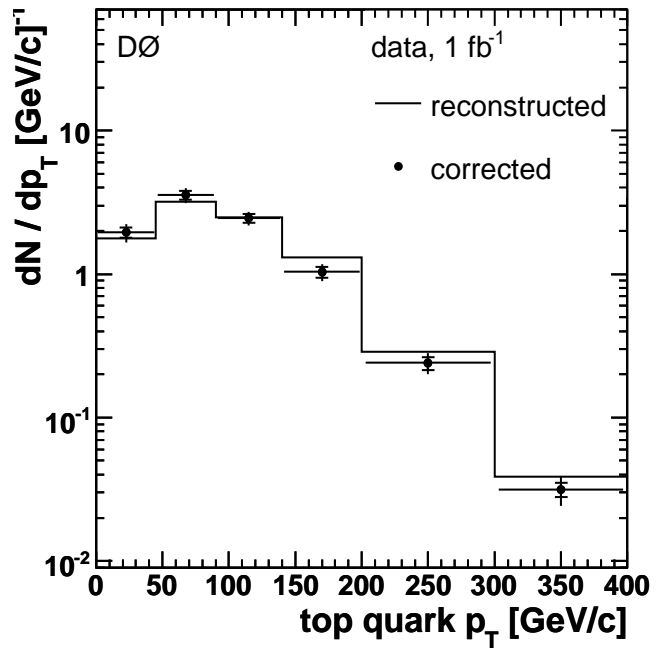}
\caption{Comparison between the background-subtracted reconstructed
  top-quark $p_T$ spectrum and the one corrected for the effects of
  finite experimental resolution (two entries per event). Inner and
  outer error bars represent the statistical and total (statistical
  and systematic added in quadrature) uncertainties, respectively.}
\label{fig:figure3}
\end{figure}

The reconstructed top-quark mass ($m_t$) from the
best fit in data, simulated $t\bar{t}$ signal, and background is shown
in~Fig.~\ref{fig:figure1}.  There is good agreement between the data
and the sum of signal and background expectations in terms of the
shape, resolution, and mean of the distribution in $m_t$ ($\chi^2/{\rm
  NDF}=1.28$).  The $p_T$ spectrum of the top quark (for leptonic and
hadronic entries) in data, together with predicted signal and
background, is shown in~Fig.~\ref{fig:figure2} for the best solution
but now refitted with a top-quark mass fixed to $170\,{\rm GeV}/c^2$ (the
value used in the inclusive cross section measurement~\cite{xsect_ref}) to
improve resolution.  To obtain a background-subtracted data spectrum,
the signal purity is fitted using signal and background contributions
as a function of $p_T$, and applied as a smooth multiplicative factor
to the data.  The result is the background-corrected distribution
shown as a solid line in Fig.~\ref{fig:figure3}.

The reconstructed $p_T$ spectrum is subsequently corrected for effects
of finite experimental resolution, based on a regularized unfolding
method~\cite{SVD_ref,GURU_ref} using a migration matrix between the
reconstructed and parton $p_T$ derived from simulation. 
The size of the $p_T$ bins was chosen based on the requirement
that the purity (the fraction of parton-level events which are 
reconstructed in the correct $p_T$ range) is $>50$\%, as 
shown in Table~\ref{table:migration}. This also
results in $p_T$ bins which are larger than the experimental
resolution for the top quark $p_T$. The correlation 
between reconstructed and correct $p_T$ is $>80$\%.
Figure~\ref{fig:figure3} compares the reconstructed and
corrected results as a function of the $p_T$ of the top
quark. 
The dependence of the unfolding on the parton spectrum shape in the
migration matrix is tested by reweighting the distribution with
arbitrary functions.  Shape variations of $\approx 20\%$ induce
$2-6\%$ changes in the differential cross section.
A correction for acceptance from the dependence of the spectrum on
kinematic restrictions of reconstructed quantities is applied to the
unfolded distributions.

\begin{figure}[t]
\includegraphics[width=0.50\textwidth,trim=1 10 10 20,clip=true]{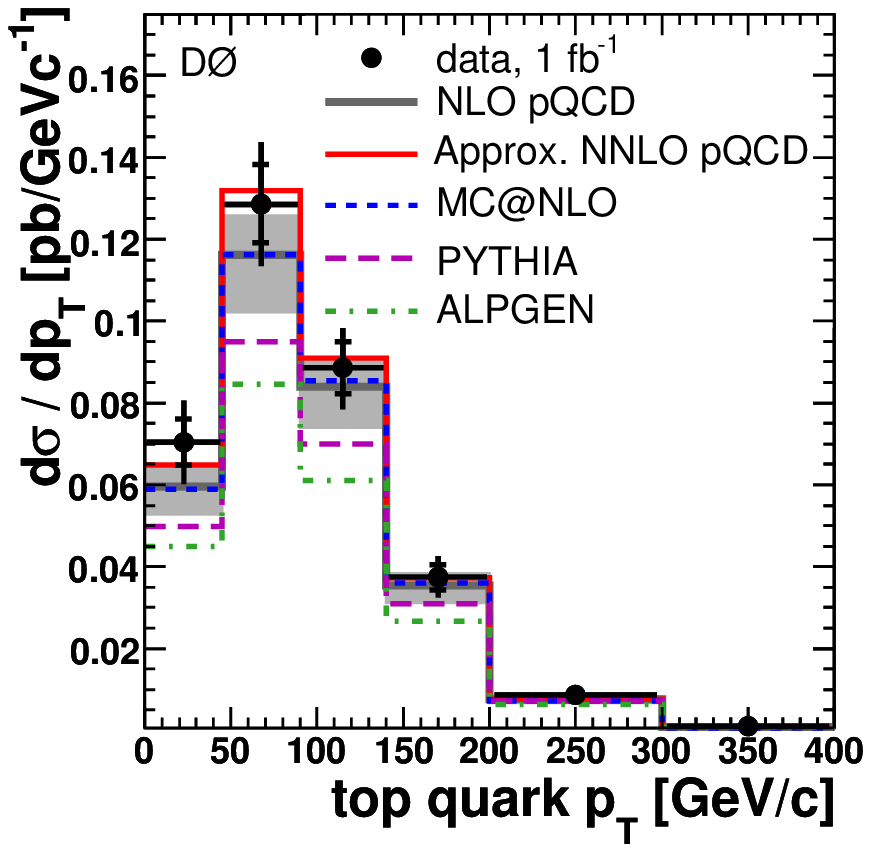}
\caption{Inclusive ${\rm d}\sigma/{\rm d}p_T$ for $t\bar{t}$
  production (two entries per event) in data (points) compared with
  expectations from NLO pQCD (solid lines), from an approximate NNLO
  pQCD calculation, and for several event generators (dashed and
  dot-dashed lines).  The gray band encompasses uncertainties on the
  pQCD scale and parton distribution functions.  Inner and outer error bars represent the statistical
  and total (statistical and systematic added in quadrature)
  uncertainties, respectively.}
\label{fig:figure4}
\end{figure}

\begin{table}[b]
  \caption{The migration matrix between the reconstructed (rows) and 
    parton (columns) top-quark $p_T$ derived from {\sc alpgen} $t\bar{t}$ events 
    passed through full detector simulation.  The matrix indicates the 
    fraction of events migrated from a given parton bin to the 
    reconstructed bins.  The binning used for 
    correlating reconstructed and parton levels of~$p_T$ are given at 
    the left and top, respectively.  Results in bold print are for 
    diagonal terms.}
\label{table:migration}
\begin{tabular}{r|llllll} \hline \hline
$p_T$ (GeV$/c)$ &  0--45 &    45--90 &   90--140 &  140--200 &  200--300 &  300--400 \\
\hline
   0--45 &     {\bf 0.530} &      0.162 &     0.062 &     0.020 &    0.003 &          0.000 \\  
  45--90  &     0.344 & {\bf 0.578} &      0.227 & 0.072 &     0.021 &          0.000 \\
  90--140 &     0.103 &   0.228 &  {\bf 0.560} & 0.223 &     0.055 &  0.031 \\
 140--200 &     0.019 &   0.029 &   0.145  & {\bf 0.581} &      0.232 &  0.071 \\
 200--300 &     0.002 &   0.002 &   0.006  &  0.103 & {\bf 0.650} &  0.363 \\
 300--400  &     0.000 &   0.000 &   0.000  &  0.001 &   0.038  &  {\bf 0.535} \\
\hline \hline
\end{tabular}
\end{table}
The measured differential cross section as a function of the $p_T$ of
the top quark (using for each event the two measurements obtained from
the leptonic and hadronic top quark decays), ${\rm d}\sigma / {\rm
  d}p_T$, is shown in Fig.~\ref{fig:figure4} and tabulated in
Table~\ref{table:xsect} together with the NLO pQCD prediction
\cite{frixione,Nason:1989zy}.
%
The statistical uncertainties are estimated by performing
1000~pseudo-experiments where, in each experiment, the
background-corrected spectrum is allowed to vary according to Poisson
statistics and is then unfolded using the regularized migration matrix
(Table~\ref{table:migration}).
The largest experimental uncertainties affecting the shape of
the $p_T$ distribution include jet energy calibration in data and in
simulation ($1.5-5.0\%$), jet reconstruction efficiency ($0.7-3.5\%$),
and jet energy resolution ($\approx0.5\%$).  
The residual dependence of the unfolded result on the top-quark mass
is $2-6\%$ for $m_{t}$ in the 170-175 GeV/$c^2$ range. This additional
uncertainty does not need to be considered for comparisons with models
in which $m_{t}$ is set to 170 GeV/$c^2$.
For the main background sources, $W/Z$+jets, we have also considered the
variations of the background shape caused by uncertainties in the
k-factors and in additional scale factors for heavy-flavour jets.
Other systematic uncertainties~\cite{xsect_ref}
account for uncertainties in the modeling of the signal, estimated
from the difference between {\sc alpgen} and {\sc pythia}, for
uncertainties in the PDFs and in the $b$-quark fragmentation.
The uncertainty on the integrated luminosity is $6.1\%$.
The systematic uncertainties quoted in the following  combine the 
uncertainty on the normalization (independent of $p_T$) with the shape-dependent
systematics.
The total correlated systematic uncertainty is $9.6\%$ (including the
uncertainty on luminosity) and the total systematic uncertainty on the
cross section, integrating over $p_T$, is $10.7\%$.

\begin{figure}[t]
\includegraphics[width=0.50\textwidth,trim=1 10 10 20,clip=true]{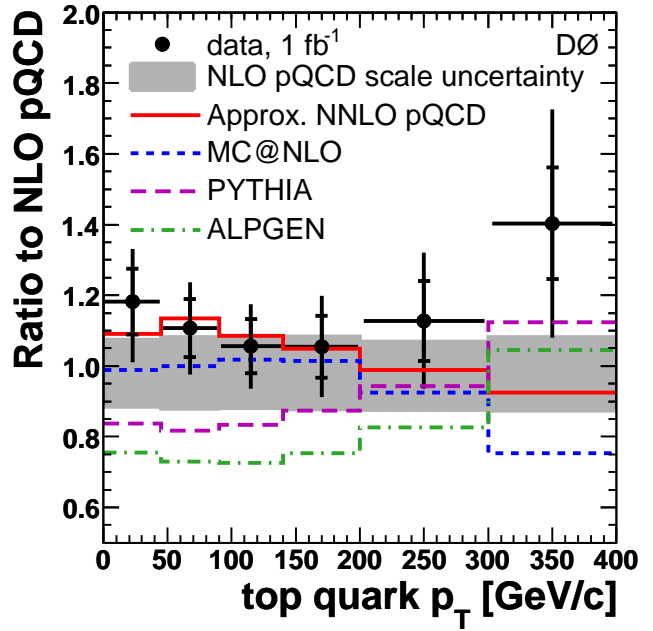}
\caption{Ratio of ${\rm d}\sigma/{\rm d}p_T$ for top quarks in
  $t\bar{t}$ production (two entries per event) to the expectation
  from NLO pQCD.  The gray band encompasses uncertainties on the scale of pQCD and parton distribution functions.
  Also shown are ratios relative to NLO pQCD for an
  approximate NNLO pQCD calculation and of predictions for several
  event generators. Inner and outer error bars represent statistical
  and total (statistical and systematic added in quadrature)
  uncertainties, respectively.}
\label{fig:figure5}
\end{figure}

\begin{figure}[!]
\includegraphics[width=0.50\textwidth,trim=1 10 10 20,clip=true]{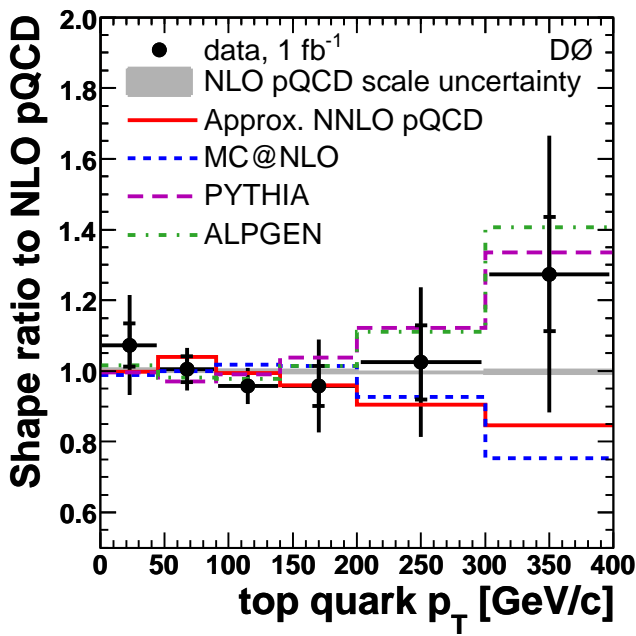}
\caption{Ratio of $(1/\sigma)\;{\rm d}\sigma/{\rm d}p_T$ for top
  quarks in $t\bar{t}$ production (two entries per event) to the
  expectation from NLO pQCD.  
  The gray band encompasses uncertainties on the scale of pQCD and parton distribution functions.
  Also shown are ratios relative to NLO
  pQCD for an approximate NNLO pQCD calculation and of predictions for
  several event generators. Inner and outer error bars represent
  statistical and total (statistical and systematic added in
  quadrature) uncertainties, respectively.}
\label{fig:figure6}
\end{figure}

Results from NLO pQCD~\cite{frixione,Nason:1989zy} calculations
obtained using CTEQ61~\cite{cteq61} PDFs
(using the scale $\mu=m_{\rm t}=170\,{\rm GeV}/c^2$) are
overlaid on the measured differential cross section in
Fig.~\ref{fig:figure4}.  Also shown are results from an approximate
next-to-NLO (NNLO) pQCD calculation~\cite{kidonakis}
computed using MSTW2008 NLO PDFs~\cite{MSTW2008} and same scales choices as
the NLO result, and from the {\sc mc@nlo}~\cite{MCatNLO} (using CTEQ61 PDFs),
{\sc alpgen}, and {\sc pythia} event generators.
The QCD scale uncertainty was evaluated for the NLO pQCD calculation
\cite{frixione,Nason:1989zy} by varying $\mu = m_{\rm t} = 170\,{\rm
GeV}/c^2$ by factors of~$2$ and $1/2$, and the PDF uncertainty by the
approximate NNLO code~\cite{kidonakis}. The total uncertainty is
$<4$\% with only a small ($<1$\%) shape variation.
A comparison of the ratio of ${\rm d}\sigma/{\rm d}p_T$
relative to a NLO pQCD calculation is shown in Fig.~\ref{fig:figure5}.  The
NLO pQCD calculations agree with the measured cross section, however,
results from {\sc alpgen} ({\sc pythia}) have a normalization shift of
about 45\% (30\%) with respect to data.  A shape comparison of the
ratio of $(1/\sigma)\;{\rm d}\sigma/{\rm d}p_T$ relative to NLO pQCD
is shown in Fig.~\ref{fig:figure6}.  All of the calculations reproduce
the observed shape.
The $\chi^2$ and corresponding $\chi^2$
probabilities~\cite{Gagunashvili:stat} for the comparisons
in~Figs.~\ref{fig:figure5} and~\ref{fig:figure6} of predictions to
data are given in~Table~\ref{table:chi2tests}.
\begin{table}[b!]
  \caption{Inclusive differential cross section ${\rm d}\sigma/{\rm d}p_{\rm T}$ 
    for $t\bar{t}$ production at $\sqrt{s}=1.96$~TeV and $m_t=170$~GeV$/c^2$.  
    There are two entries per event, with the total normalized to the $t\bar{t}$ production cross section.  
    In addition to total systematic uncertainties on the shape in $p_T$ in each bin, 
    there is a $p_T$-independent systematic uncertainty of $9.6\%$ that is not 
    included in the table.
  }
\label{table:xsect}
{\begin{tabular}{D{-}{-}{3}D{.}{.}{1}dddd}
\hline\hline
\multicolumn{1}{c}{\multirow{2}{*}{$p_T$}} &  \multicolumn{1}{c}{\multirow{2}{*}{\text{$\langle p_T\rangle$}}} &  \multicolumn{1}{c}{\text{Cross}} &  \multicolumn{1}{c}{\text{Stat.}}  &  \multicolumn{1}{c}{\text{Shape Sys.}}  &   \multicolumn{1}{c}{\text{NLO}}  \\ 
       &       &  \multicolumn{1}{c}{\text{Section}} &   \multicolumn{1}{c}{\text{Unc.}}   &  \multicolumn{1}{c}{\text{Unc.}}  &  \multicolumn{1}{c}{\text{pQCD}}  \\ 
\multicolumn{1}{c}{(GeV$\!/c$)} & \multicolumn{1}{c}{\text{(GeV$\!/c$)}} & \multicolumn{1}{c}{\text{(fb/GeV)}} & \multicolumn{1}{c}{\text{(fb/GeV)}} & \multicolumn{1}{c}{\text{(fb/GeV)}} & \multicolumn{1}{c}{\text{(fb/GeV)}} \\ 
\hline
     0-45 &    29 &     70 & 11 & 5 & 59.6 \\  
    45-90 &       68 &      130 & 20 & 10 & 116 \\  
   90-140 &       113 &     89 & 13 & 6 & 83.8 \\  
  140-200 &       165 &     37 & 6 & 3 & 35.6 \\  
  200-300 &       233 &   8.7 & 1.7 & 0.7 & 7.72 \\  
  300-400 &       329 &   1.1 & 0.3 & 0.1 & 0.814 \\ \hline 
\multispan{2}{$\sigma_{t\bar{t}}$ (pb)\hfill} & 8.31 & 1.28 &      & 7.54 \\  
\hline\hline
\end{tabular}
}
\end{table}
\begin{table}[b]
  \caption{The $\chi^2/$NDF and $\chi^2$ probability for comparisons 
    between the measured data and predictions using 
    correlated (uncorrelated) uncertainties for the absolute (shape) comparison.}
\label{table:chi2tests}
\begin{tabular}{lllll}
\hline\hline
\multirow{2}{*}{Prediction}  & \multicolumn{2}{c}{Absolute\hspace*{0.8cm}} & \multicolumn{2}{c}{Shape} \\ 
  &  $\chi^2\!/$NDF  & prob. &  $\chi^2\!/$NDF  & prob. \\ \hline  
NLO pQCD           & 0.695 & 0.653 & 0.315 & 0.904 \\
Approx. NNLO pQCD  & 0.521 & 0.793 & 0.497 & 0.779 \\
{\sc mc@nlo}         & 1.22 & 0.295 & 0.777 & 0.566 \\
{\sc pythia}         & 2.61 & 0.0157 & 0.352 & 0.881 \\
{\sc alpgen}         & 5.04 & 3.54$\times 10^{-5}$ & 0.204 & 0.961 \\
\hline\hline
\end{tabular}
\end{table}

In conclusion, we have presented a $1\,{\rm fb}^{-1}$ measurement of
the differential cross section of the top-quark $p_T$ for $t\bar{t}$
production in the $\ell+$jets channel using $p\bar{p}$ collisions at
$\sqrt{s}=1.96$~TeV.
Results from NLO and NNLO pQCD calculations and from the {\sc mc@nlo}
event generator agree with the normalization and shape of the measured
cross section.  Results from {\sc alpgen+pythia} and {\sc pythia}
describe the shape of the data distribution, but not its
normalization.
%
We thank the staffs at Fermilab and collaborating institutions, 
and acknowledge support from the 
DOE and NSF (USA);
CEA and CNRS/IN2P3 (France);
FASI, Rosatom and RFBR (Russia);
CNPq, FAPERJ, FAPESP and FUNDUNESP (Brazil);
DAE and DST (India);
Colciencias (Colombia);
CONACyT (Mexico);
KRF and KOSEF (Korea);
CONICET and UBACyT (Argentina);
FOM (The Netherlands);
STFC and the Royal Society (United Kingdom);
MSMT and GACR (Czech Republic);
CRC Program, CFI, NSERC and WestGrid Project (Canada);
BMBF and DFG (Germany);
SFI (Ireland);
The Swedish Research Council (Sweden);
and
CAS and CNSF (China).
%

\vfill\eject


\begin{thebibliography}{99}
\bibitem{eta} Pseudorapidity is defined as $\eta =
  -\ln\tan(\theta/2)$ where $\theta$ is the angle measured with
  respect to the proton beam coinciding with the positive $z$ axis of
  a right-handed coordinate system at the center of the detector.
\end{thebibliography}
\end{document}